\documentclass[12pt,preprint]{aastex}

\def\apjs{{\rm ApJS}}                  
\def\apjl{{\rm ApJ Let}}               
\def\pasp{{\rm PASP}}                  

\def\Msol{\thinspace\hbox{$\hbox{M}_{\odot}$}}

\def\a4{\hsize 17.0cm \vsize 25.cm}
\newcommand{\der}[2]  { \frac{{\rm d}#1}{{\rm d}#2} }

\newcommand{\dif}     {{\rm d}}

\shorttitle{On the  Constrains to the SSCs Heating Efficiency}
\shortauthors{Silich et al.}

\begin{document}

\title{On the Heating Efficiency Derived from Observations of 
       Young Super Star Clusters in M82}

\author{ 
Sergiy Silich\altaffilmark{1}, Guillermo Tenorio-Tagle\altaffilmark{1}, 
Ana Torres Campos\altaffilmark{1},} 
\author{Casiana Mu\~noz-Tu\~n\'on\altaffilmark{2}, 
Ana Monreal-Ibero\altaffilmark{2,3}, Veronica Melo\altaffilmark{2}}
\altaffiltext{1}{Instituto Nacional de Astrof\'\i sica Optica y
Electr\'onica, AP 51, 72000 Puebla, M\'exico; silich@inaoep.mx}
\altaffiltext{2}{Instituto de Astrof\'{\i}sica de Canarias, E 38200 La
Laguna, Tenerife, Spain; cmt@ll.iac.es}
\altaffiltext{3}{European Southern Observatory, Karl-Schwarzschild-Strasse 2
D-85748 Garching bei M\"unchen, Germany}

\begin{abstract}

Here we discuss the mechanical feedback that massive stellar clusters 
provide to the interstellar medium of their host galaxy.
We apply an analytic theory developed in a previous study for M82-A1 to a 
sample of 10 clusters located in the central zone of the starburst galaxy M82,
all surrounded by compact and dense HII regions. We claim that the only 
way that such HII regions can survive around the selected clusters, is if 
they are embedded into a high pressure ISM and if the majority of their 
mechanical energy is lost within the star cluster volume via strong 
radiative cooling. The latter implies that these clusters have a low heating 
efficiency, $\eta$, and evolve in the bimodal hydrodynamic regime. In this 
regime the shock-heated plasma in the central zones of a cluster 
becomes thermally unstable, loses its pressure and is accumulated 
there, whereas the matter injected by supernovae and stellar winds outside 
of this volume forms a high velocity outflow - the star cluster wind. 
We calculated the heating efficiency for each of the selected clusters and
found that in all cases it does not exceed 10\% . Such low heating efficiency 
values imply a low mechanical energy output and the impact that the selected 
clusters provide to the ISM of M82 is thus much smaller than what one would 
expect using stellar cluster synthetic models.
\end{abstract}

\keywords{galaxies: individual (M82) --- galaxies: star clusters --- HII 
regions --- ISM: bubbles --- kinematics and dynamics}

\section{Introduction}

It is a common believe that massive star clusters return a  significant 
fraction of their stellar mass to the interstellar medium (ISM). This is 
thought to be done in a violent manner  that deeply affects the structure of 
the interstellar gas and in the case of starburst galaxies, this may even 
result  in  the channeling  of the processed material into the
intergalactic space (see, for instance,  Tenorio-Tagle et al. 2003, 
Cooper et al. 2008, and references therein). 
The  general concensus is that within the volume occupied by superstar 
clusters (SSCs), the kinetic energy supplied by massive stars in the form of 
stellar 
winds and supernovae explosions is there in situ thermalized. This results 
into a high temperature ($T \sim 10^7$K) plasma, with a large thermal pressure
that highly exceeds that in the ambient ISM, and this provokes the exit of 
the thermalized ejecta  out of the cluster as a supersonic star cluster wind 
(Chevalier \& Clegg, 1985). The cluster winds shape the ISM  by generating 
large-scale superbubbles. These shock and displace the surrounding ISM while 
locking it into large expanding shells able to cool down by radiation in a 
short characteristic time scale, while the much lower density shock-heated 
wind gas, which fills the superbubble interior, remains hot for a considerably
longer time (Weaver et al. 1977;  Mac Low \& McCray, 1988; Tenorio-Tagle et 
al. 2006) and promotes the 
growth of the superbubble. This shocked wind plasma has been  detected around 
OB-associations and stellar clusters as a soft X-ray emitter (see, for 
example, Chu et al. 1995; Stevens \& Hatwell 2003, Silich et al. 2005 and
references therein), whereas 
the outer shells have been traced in 21 cm (Puche et al. 1992, Ehlerov\'a 
et al. 2004) or as photoionized engulfing 
filaments, if in presence of a strong  Lyman continuum radiation (Meaburn, 
1980;  Lozinskaya, 1992 and references therein).
The size and interior pressure of superbubbles in the case 
of an homogeneous interstellar gas distribution  (see Mac Low \& McCray, 1988;
Bisnovatyi-Kogan \& Silich 1995 and references therein) are: 
\begin{eqnarray}
      \label{eq1a}
      & & \hspace{-1.0cm}
R_{sb} = \left(\frac{375(\gamma-1)}{28(9\gamma-4)\pi}\right)^{1/5} 
         \left(\frac{L_{out}}{\rho_{ISM}}\right)^{1/5} t^{3/5} , 
      \\[0.2cm]
      \label{eq1.c}
      & & \hspace{-1.0cm}
P_{sb} = 7 \rho_{ISM} \left[\frac{3(\gamma - 1)}{700 (9\gamma - 4) \pi}
         \frac{L_{out}}{\rho_{ISM}}\right]^{2/5} t^{-4/5} ,
\end{eqnarray}  
where $\gamma = 5/3$ is the ratio of specific heats, $R_{sb}$ 
is the outer shell radius, 
$P_{sb}$ is the pressure in the shock-heated wind region, 
$L_{out}$ is the star cluster mechanical energy output, $\rho_{ISM}$ is the 
interstellar gas density, and $t$ the evolutionary time. 
The superbubbles are supposed to expand until they acquire pressure 
equilibrium with the surrounding medium ($P_{sb} = P_{ISM}$), when:
\begin{equation}
      \label{eq1d}
R_{sb} = (7 \gamma)^{3/4} 
         \left[\frac{3 (\gamma - 1)}{28 (9\gamma - 4) \pi}
         \frac{L_{out}}{\rho_{ISM} a^3_{ISM}}\right]^{1/2} = 
       637 \left(\frac{L_{38}}{n_{ISM} a^3_{10}}\right)^{1/2} \, pc  , 
\end{equation}  
where $n_{ISM}$ is the interstellar gas number density, $L_{38}$ is the
mechanical energy output in units of $10^{38}$~erg s$^{-1}$ and $a_{10}$
is the sound speed in the interstellar medium in units of 10~km s$^{-1}$.

However, as noticed by Silich et al. (2007), this cannot be the whole story.
The observed properties of the HII region associated to the massive 
(1.3$ \times 10^6$ M$_\odot$), young (age $\sim$ 6 Myr) and thus powerful 
(with a mechanical luminosity, $L_{mech} \approx 2.5 \times 10^{40}$ erg 
s$^{-1}$) super star cluster M82-A1 (Smith et al. 2006) are not consistent 
with the interstellar bubble model (equations \ref{eq1a} - \ref{eq1d}). The 
associated minute, low mass ($M_{HII} \approx 5 \times 10^3$~M$_\odot$),
although dense ($n_{HII} \approx 1800$~cm$^{-3}$), HII region presents a  
radius ($R_{HII} \approx 4.5$~pc) much smaller than that predicted by 
equations (\ref{eq1a} and \ref{eq1d}). It seems surprising that M82-A1
and other young and massive clusters in M82 (Melo et al. 2005), NGC 3351 
(H\"agele et al. 2007) and in other galaxies are surrounded by compact, low 
mass (see Table 2 below) HII regions despite the powerful mechanical energy 
output predicted for the clusters by stellar evolution synthesis models. 

Silich et al. (2007) suggested that in this case clearly only a fraction of 
the star cluster mechanical luminosity is converted into the energy of the 
outflowing plasma whereas the rest ought to be lost due to strong 
radiative cooling. They also developed an analytic and semi-analytic model, 
which led to obtain the value of the heating efficiency $\eta$, the parameter 
which links the star cluster mechanical luminosity with the actual thermal 
energy that is deposited into the star cluster volume. The models reveal the 
value of the heating efficiency by fitting the ionized gas number density and 
radius of the compact HII region detected around a massive star cluster. The 
results led to a low heating efficiency ($\eta < 10$\%) in the case of M82-A1.

Here we extend the analysis of Silich et al. (2007) to a sample of 10 SSCs 
selected from the list of Melo et al. (2005), in order to reveal their
heating efficiency, and thus the energy that these clusters return to the ISM
of M82.

Section 2 and 3 discuss the hydrodynamics of SSCs, the model assumptions and 
present the equations used in order to obtain the heating efficiency. The 
sample of the selected clusters is presented in section 4. We apply our model 
to each of the selected clusters and discuss our results in section 5.

\section{The heating efficiency in SSCs}

The hydrodynamics of the matter returned by stellar winds and supernova 
explosions within the cluster volume has been approximated assuming first 
that the sources are equally spaced within a spherical volume of radius
$R_{SC}$.  In the pioneer adiabatic approach of Chevalier \& Clegg (1985), the 
kinetic energy supplied by the evolving massive stars, $L_{mech}$, has been 
assumed to be  completely converted into thermal energy of the hot plasma. 
The strong pressure gradient generated by the deposited matter, forces then 
the gas velocity to increase almost linearly from $0$~km s$^{-1}$ at the star
cluster center to its sound speed at the cluster edge. Once the gas streams out 
of the cluster, it rapidly acquires its terminal speed ($v_{A\infty} 
\sim 2 c_{SC}$), while its density and temperature drop as $r^{-2}$, and 
$r^{-4/3}$, respectively.
   
More recently, Silich et al. (2004), Tenorio-Tagle et al. (2005, 2007) and 
W\"unsch et al. (2008) recognized that the adiabatic assumption is not 
valid in the case of massive and compact star clusters and developed a 
radiative star cluster wind model, which takes into consideration the energy 
losses  that occur in the hot thermalized plasma. They found a 
threshold line, $L_{crit}(R_{SC})$ in the $L_{mech} - R_{SC}$ parameter space.
The radiative solution is in excellent agreement with Chevalier \& Clegg's 
results in the case of low mass clusters, when $L_{mech} \ll L_{crit}$. 
However, strong radiative cooling modifies essentially the temperature 
distribution outside of the cluster when the star cluster mechanical
luminosity approaches the threshold value, $L_{mech} \le L_{crit}$.
When the mechanical luminosity of the considered clusters exceeds the 
threshold value, $L_{crit}$, catastrophic cooling sets in within the central
zones of the cluster, what results into a bimodal flow regime. In this case 
the stagnation radius, $R_{st}$, moves out 
of the cluster center and splits the cluster volume into two distinct zones. 
In the inner zone, $r < R_{st}$, strong radiative cooling promotes frequent 
thermal instabilities in the injected gas, reducing significantly  the 
pressure gradient and thus the outward acceleration. Strong radiative cooling 
thus leads to the accumulation of the matter injected within  the volume 
defined by the stagnation surface.  In the outer zone, 
$R_{st} < r  < R_{SC}$, despite radiative cooling, the pressure gradient 
remains  sufficient  to drive the injected matter away from  the cluster, as 
a strongly radiative stationary wind. 

The detail physics  during  the thermalization process are however not well 
understood. In the original paper of Chevalier \& Clegg (1985)
it was assumed that the amount of the deposited thermal 
energy per unit volume, $q_{e}$, is identical to the rate of mechanical 
energy released by massive stars: $q_{e} = q_{mech}$. However Stevens \& 
Hartwell (2003) found that this assumption is not in good agreement with
the spectra of the diffuse X-ray emission detected in a number of nearby 
massive clusters. Bradamante et al. (1998), and Recchi et al. (2001), who 
studied the chemical and dynamical evolution of blue compact galaxies, also 
claimed that only a few per cent of the energy deposited by supernovae type II 
provides the energetics  of the host galaxy ISM, while the rest is 
radiated away. It is therefore highly desirable to link the value of 
the heating efficiency with stellar clusters observable quantities.

A firm evidence for an incomplete transformation of the star 
cluster mechanical luminosity into the energy of the star cluster wind was
obtained by Smith et al. (2006), who provided detailed photometric and
spectral analysis of the massive, young SSC M82-A1 and its associated
HII region. This led them, as well as to Silich et al. (2007), to claim 
that the energy $q_{e}$ represents only a small fraction of the mechanical 
energy provided by massive stars: 
$q_{e} = \eta q_{mech}$, with $\eta << 1$. The physical justification for this
parameter comes from the fact that strong radiative cooling may  take place 
during the process of thermalization either because of an enhanced  gas 
metallicity, resultant from SN explosions, or because of the large densities 
within the shock-heated zones between neighboring  massive stars, before the 
newly injected matter joins the flow (W\"unsch et al. 2007; Silich et al. 
2007). In this case only a fraction of the mechanical energy 
supplied by the collection of massive stars is shared by the matter within 
the cluster volume and thus the actual thermal energy given to the injected 
gas is smaller than that provided by the collection of  massive stars. This 
is particularly important in the case of massive ($M_{SC} \ge 10^5$ M$_\odot$)
and compact ($R_{SC} \sim$ a few parsecs) clusters which present  a large 
massive star number density, $N_{\star}$, and a small mean separation between 
them: $\Delta R = N^{-3}_{\star} \ll 1$~pc.

Indeed, the multiple interactions expected between supersonic stellar winds 
and the supernovae ejecta in such compact and massive clusters are similar to 
those occurring 
in colliding wind binaries that lead to a  shock-heated plasma, that
effectively radiates in the soft X-ray regime. Luo et al. (1990); 
Stevens et al. (1992) found that in the case of colliding wind binaries 
the amount of energy radiated away from the shock-heated zone, $L_{lost}$, 
depends on the binary separation. It scales as $L_{lost} \sim \Delta R^{-1}$ 
in the quasi-adiabatic regime and increases when radiative cooling in the 
shock-heated zone is taken into consideration.
In the case of a dense stellar cluster the kinetic energy 
placed by massive stars interacts with that deposited by multiple nearby 
neighbors. This suggests that the energy, which actually
drives the star cluster outflow, is smaller than the total provided by the 
massive stars within the cluster volume, particularly if one accounts for the 
large metallicities expected from supernovae. 

In the semi-analytic models all uncertainties dealing with the distribution of
massive stars and the collisions between nearby supersonic flows and thus the
sudden loss of energy, are accounted for by the parameter $\eta$, known as 
the heating efficiency. The fraction of energy that a star cluster 
returns to the ambient interstellar gas strongly depends on this parameter. 
Thus $\eta$ defines the mechanical feedback that star clusters provide to the 
ISM of their host galaxy.

\section{The pressure confined wind model}

The pressure confined wind model (Silich et al. 2007) suggests that the 
combination of two factors is crucial in order to produce the compact and 
dense HII regions able to  survive around powerful young clusters. 
These are a low heating efficiency and a large thermal pressure, $P_{ISM}$, 
in the surrounding ISM, what leads to a pressure confined bubble 
configuration. 
Thus in this model the size of the standing HII region depends critically 
on the balance between $P_{ISM}$ and the wind ram pressure at the reverse 
shock position ($P_{ram} = P_{ISM}$). In this case the structure of 
the outflow can be derived 
analytically from a set of equations that consider:  conservation of mass, 
photoionization balance, pressure equilibrium and the fast radiative cooling 
that occurs within  the star cluster volume and on its wind. Our set of 
equations is such that if the parameters of the driving stellar cluster: 
its mass ($M_{SC}$), radius ($R_{SC}$), the number of ionizing photons 
($N^{SC}$) are known, one can match the model predicted radius ($R_{HII}$) 
and gas number density ($n_{HII}$) of the associated HII region with the 
observed values. In this approach, one can find the value of the heating 
efficiency from a nonlinear algebraic equation which relates $\eta$ with the 
host cluster and the associated HII region parameters (see Silich et al. 2007):
\begin{equation}
      \label{eq2}
1 - \frac{(4 \pi P_{ISM} V^2_{A\infty} R^2_{HII})^{1/2}}
    {(4 L_{crit} L_{SC} V^2_{\infty})^{1/4}} 
\left[\left(1 - \frac{3 f_t N^{SC}}
{4 \pi \beta n^2_{HII} R^3_{HII}}\right)^{1/3}
      - \frac{9}{512} \frac{f_{\lambda} \mu^2_i V^5_{\infty}}
        {P_{ISM} R_{HII} \Lambda_{s}}\right] = 0 ,
\end{equation}
where $V_{A\infty} = (2 L_{SC} / {\dot M}_{SC})^{1/2}$ is 
the adiabatic wind terminal speed, $L_{SC}$ and ${\dot M}_{SC}$ are the star 
cluster mechanical luminosity and the mass input rate, respectively,
$\Lambda_{s}$ is the value of the cooling function at the reverse shock 
radius, $\beta$ is the recombination coefficient to all but the ground level, 
$f_{\lambda} = 0.3$ is a fiducial coefficient and $f_t$ is the fraction of 
ionizing photons, which reaches the outer standing gaseous shell. 
Note that equation (\ref{eq2}) is only valid in the bimodal parameter space, 
i.e. it can only be applied to clusters with a mechanical power that exceeds 
the threshold value $L_{crit}$, and carries a strong implicit dependence on 
$\eta$  via the threshold mechanical luminosity, $L_{crit}$ and the star 
cluster wind terminal speed, $V_{\infty}$ (see W\"unsch et al. 2007; 
Silich et al. 2007):
\begin{eqnarray}
\label{eq2a}
      & & \hspace{0.0cm}
L_{crit} = \frac{3 \pi \eta \alpha^2 \mu_i^2 R_{SC} V^4_{A\infty}}
      {2 \Lambda_{st}}
      \left(\frac{\eta V^2_{A\infty}}{2}  - \frac{c^2_{st}}{\gamma-1}\right) ,
      \\[0.2cm]
      \label{eq2b}
      & & \hspace{0.0cm}
V_{\infty} = [2/(\gamma-1)]^{1/2} c_{st} 
\end{eqnarray}
where  $\alpha = 0.28$ and $\mu_i = 14 m_H / 11$ is the mean mass per ion.
$\Lambda_{st}$ and $c_{st}$ are the values of the cooling function and the
speed of sound at the stagnation point, both are functions of temperature 
at the stagnation radius, $T_{st}$, which strongly depends also on $\eta$. 
We obtain the value of $T_{st}$ from the condition that the stagnation 
pressure reaches the maximum possible value and thus 
$\dif{P_{st}}/\dif{T_{st}} = 0$ (Tenorio-Tagle et al. 2007): 
\begin{equation}
      \label{eq3}
\left(\frac{\eta V^2_{A\infty}}{2} - \frac{c^2_{st}}{\gamma - 1}\right)
\left(1 - \frac{T_{st}}{2 \Lambda} \der{\Lambda}{T_{st}}\right) -
\frac{1}{2} \frac{c^2_{st}}{\gamma - 1} = 0 .
\end{equation}
Note that the calculated heating efficiency does not depend significantly on 
the parameter $f_t$. Hereafter we shall assume that $f_t = 0.5$ and 
$V_{A\infty} = 1000$~km s$^{-1}$. 

In order to relate the star cluster mass with the star cluster mechanical 
luminosity we use a relation, which approximates the results of Starburst 
99 synthesis model for coeval clusters with a Salpeter initial mass function 
with sources between 1\Msol \, and 100\Msol \, and ages in the range 
$\sim$ 4 Myr - 12 Myr (Leitherer et al. 1999):
\begin{equation}
\label{eq5}
L_{SC} = 3 \times 10^{40} \left(\frac{M_{SC}}{10^6\Msol}\right) \, erg \, 
s^{-1} .
\end{equation}
Equation (\ref{eq2}) presents only a weak dependence on $L_{SC}$ and thus 
deviations of the star cluster mechanical luminosity from the assumed 
constant value do not affect the final results significantly.

Thus, in this approach one can obtain the heating efficiency $\eta$ directly
from the observed parameters of the stellar cluster and its associated HII 
region: $M_{SC}$, $R_{SC}$, $N^{SC}$, $R_{HII}$ and $n_{HII}$ by solving 
equation (\ref{eq2}).

\section{A sample of clusters in M82}

In order to learn how efficient the conversion of the star cluster mechanical
luminosity  into the wind driving energy is, one needs a sample of clusters 
whose masses, sizes and Lyman continuum radiation are known together with 
the radius and density of their adjacent HII regions. Most of these 
parameters can be obtained from the photometric sample of Melo et al. (2005) 
who cataloged 197 young super stellar clusters in the central zone of the 
galaxy 
M82. The only parameter, which is required by equation (\ref{eq2}) and which 
Melo et al. (2005) did not obtain, is the density of the ionized gas in the 
HII regions. We obtain this quantity from  
\emph{Potsdam Multi Aperture Spectrophotometer}, PMAS (Roth et al. 2005) 
observations at the 3.5~m telescope in Calar Alto. PMAS is a very 
versatile instrument, with several working modes. Here, we used its lens 
array (LARR) which is made out of $16 \times 16$ square elements. We 
observed the nuclear region of M82 using two continuous fields with the 
spatial sampling of $0\farcs5 \times 0\farcs5$ and thus, covering a field 
of view of $8\farcs0 \times 8\farcs$0 per pointing.
Two different sets of observations were provided. A set of data with low 
spectral resolution, covering the whole optical spectral range was obtained 
in service mode on the 3rd and 4th of June 2005 while that with a high 
spectral resolution on the 2nd of February 2005. Both sets of data were taken 
under non-photometric conditions. Seeing ranged typically between 1\farcs3 
and 1\farcs6.  Line profiles were fitted 
using gaussian functions. This procedure was done in an automatic way using 
the IDL based routine MPFITEXPR implemented by Markwardt\footnote{See
http://cow.physics.wisc.edu/\~craigm/idl/idl.html.} checking each
fit, afterwards. A single gaussian fit was enough in order to reproduce 
the observed profiles. For each set of lines, wavelenght differences between 
them were fixed and the same line width was assumed. Then the intensity maps 
of $H_{\alpha}$ and $S_{II}$ lines were produced. 

Our field of view includes region A (O'Connell \& Mangano, 1978) of the 
central zone as well as a highly extincted heart-shaped region towards the 
west. In order to localize the selected clusters in the PMAS map, the 
resolution of the HST image was degraded to $0\farcs506$ / pixel, which is 
almost identical to that of PMAS ($0\farcs5$ / pixel) and a new, low 
resolution HST $H_{\alpha}$ map of M82 nuclear region was generated. 
As we did not have absolute astrometry and in order to use the  two 
observing data sets, the low-resolution HST and the PMAS $H_{\alpha}$ maps 
were compared and displaced until reaching the highest cross  correlation 
coefficient (Russ, 2002).
Then the electron density, $n_{HII}$, was derived (see Table 2) from a map of 
the \textsc{[S\,ii]}$\lambda$6717 / \textsc{[S\,ii]}$\lambda$6731 line ratio 
using the task \texttt{temden}, based on the \texttt{fivel} program (Shaw \& 
Dufour, 1995), included in the IRAF package \texttt{nebular} and assuming
an electronic temperature of 10000~K (see Figure 1).
\begin{figure}[!htbp]
\plotone{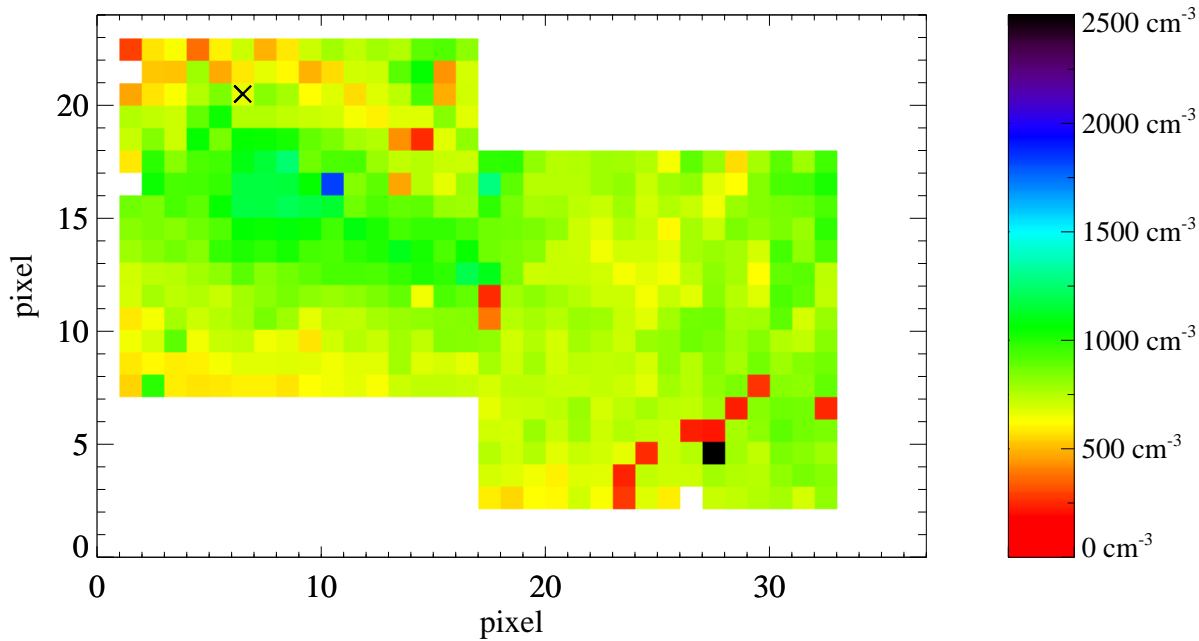}
\caption{The ionized gas density distribution derived from the PMAS 
observations. The range of densities is coded by color scale on the
right of the panel. Positions of different points in the field of
view of PMAS are given in pixels. The position of the M82-A1 cluster
is marked with a cross symbol.}
\label{fig1}
\end{figure}

We use the electron density values and the photoionization balance equation 
in order to estimate the masses of the associated HII regions: 
\begin{equation}
      \label{eqx}
M_{HII} = \frac{\mu_i N^{SC}}{\beta n_{HII}} ,
\end{equation}  
where $\mu_i$ is the mean mass per ion and $\beta  = 2.59 \times 10^{-13}$ 
cm$^{-3}$ s$^{-1}$ is the recombination coefficient to all but the ground 
level.

84 out of the 197 clusters found by Melo et al. (2005) are located in the area
observed with PMAS. 
From the large sample of young SSCs cataloged by Melo et al. (2005) we 
have selected a subsample  which follows the criteria that the radius of the 
HII region (the one  defined in the $H_{\alpha}$ HST images) lies 
clearly outside the volume ocupied by the SSCs themselves (radius taken from 
the continuum HST  images). In this way we selected a total of 21 objects. 
We then compared our list of cluster candidates with that
of Mayya et al. (2008), who used the HST Advanced Camera for Surveys (ACS)
and selected only those sources, which were simultaneously detected in 
three different (B, V and I) filters. Only 10 counterparts for our 21 
candidate clusters were found in the list of Mayya et al. (2008). 
We have selected these as genuine clusters for our further discussion. 
Figure 2 presents the location of the selected clusters 
within the galaxy and also outlines the area in the central zone of M82,
which was observed with PMAS.
\begin{figure}[!htbp]
\plotone{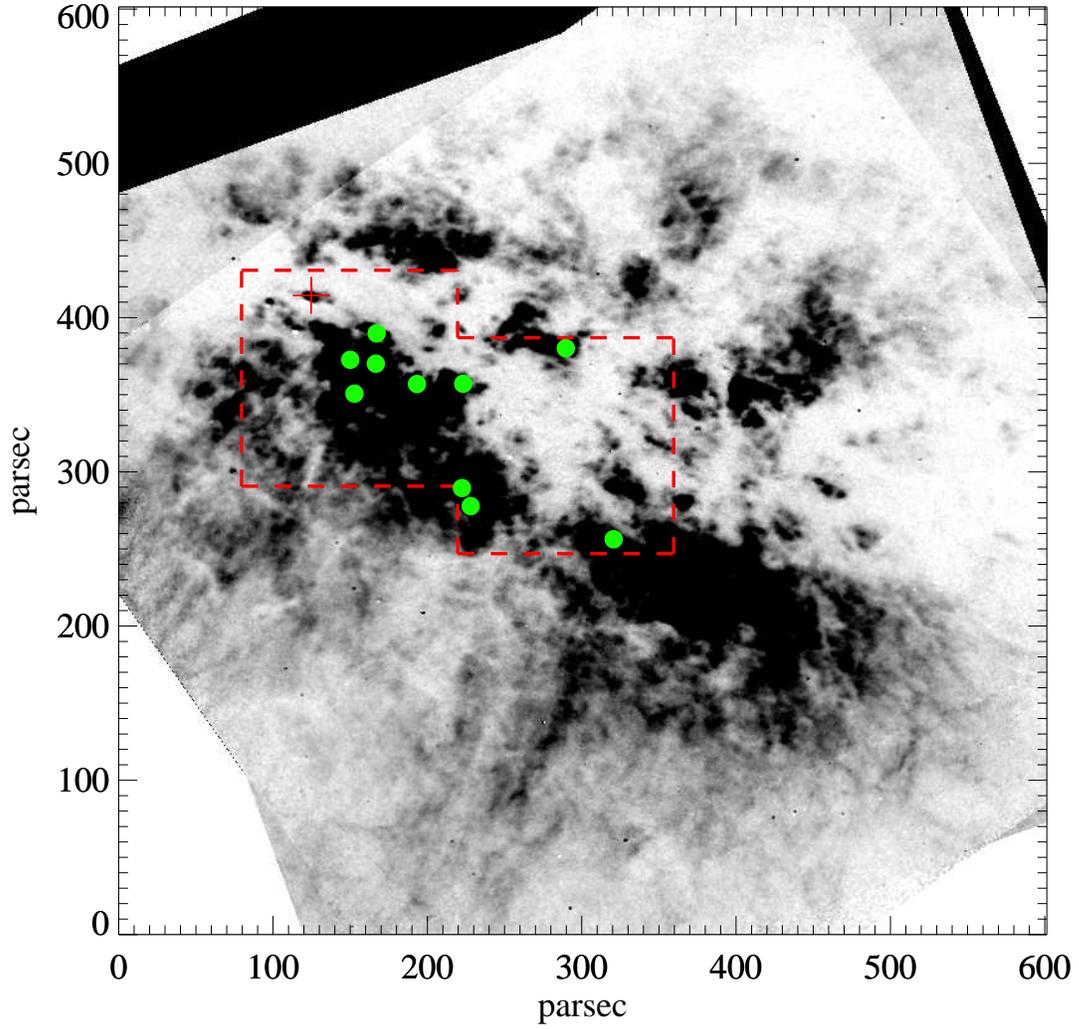}
\caption{The HST WFPC2 image of the central zone of M82 taken in the F656N 
filter. The selected clusters are shown as green dots. The position of the 
M82-A1 cluster is used as a reference point and is marked with a red plus. 
Dashed lines outline the field of view of PMAS.}
\label{fig2}
\end{figure}

Table 1 presents the identification of the selected clusters. Here the first 
column marks the clusters in our list, columns 2 and 3 provide the star 
cluster and the M82 zone identification in the sample of Melo et al. (2005) 
and column 4 lists the identification number in the sample of Mayya et al. 
(2008). 
\begin{table}[htp]
\caption{\label{tab1} Selected clusters}
{\small
\begin{tabular}{c c c c}
\hline\hline 
Cluster & ID          & zone & ID \\
        & Melo et al. (2005) &      & Mayya et al. (2008) \\
\scriptsize{(1)} & \scriptsize{(2)} & \scriptsize{(3)} &\scriptsize{(4)} \\
\hline
1&  20	& SW &  178N \\
2&  52	& SE &  47N  \\
3&  12	& SE &  13N  \\
4&  58	& SE &  34N  \\
5& 14	& SE &  20N  \\ 
6& 59	& SE &  45N  \\ 
7& 3	& NE &  71N  \\
8& 72	& SE &  200N \\
9& 81	& SE &  32N  \\ 
10& 75	& SE &  28N  \\
\hline\hline
\end{tabular}
     } 
\end{table}

Table 2 presents the star cluster masses and radii (columns 2 and 3, 
respectevely) and the number of Lyman continuum photons (column 4) taken 
from Melo et al. (2005).
The number densities, radii and masses of the associated HII regions are 
given in columns 5, 6 and 7, respectively. Column 8 presents the calculated 
values of the heating efficiency and column 9 - the output 
mechanical luminosity normalized to the star cluster mechanical luminosity,
$L_{SC}$, predicted by the Starburst 99 synthetic model.

Note that the cluster radii fall into a narrow size interval, $2 < R_{SC} < 
6$~pc whereas their masses vary from $2 \times 10^4\Msol$ to $8 \times 
10^5\Msol$. In all selected cases the resulting masses of the associated HII 
regions do not exceed a few thousand solar masses, just as 
in the case of M82-A1 whose stellar mass is $\sim 10^6$\Msol \, and its 
associated HII region has only $\sim 5000$\Msol \, (Smith et al. 2006).
\begin{table}[htp]
\caption{\label{tab1} Parameters of the selected clusters and their HII
regions}
{\small
\begin{flushleft}
\begin{tabular}{c c c c c c c c c}
\hline\hline
Cluster & M$_{SC}$ & R$_{SC}$ & N$^{SC}$
& n$_{HII}$ & R$_{HII}$ & M$_{HII}$ & $\eta$ &L$_{out}$/L$_{SC}$ \\
& {\scriptsize{($10^{5}$ M$_{\odot}$)}} & {\scriptsize{(pc)}}
& {\scriptsize{($10^{49}$ s$^{-1}$)}} & {\scriptsize{(cm$^{-3}$)}}
& {\scriptsize{(pc)}} & {\scriptsize{(M$_{\odot}$)}} & $\%$ & $\%$  \\
\scriptsize{(1)} & \scriptsize{(2)} & \scriptsize{(3)} &\scriptsize{(4)}
&\scriptsize{(5)} &\scriptsize{(6)} &\scriptsize{(7)} &\scriptsize{(8)}
&\scriptsize{(9)} \\
\hline
1  & 0.35$^{\pm 0.16}$ & 4.03 & 7.1$^{\pm 3.7}$ & 769$^{\pm 76}$ &
 5.64  & 381.6& 7.8$^{\pm 2.1}$  & 3.17$^{\pm 1.06}$ \\
2  & 0.40$^{\pm 0.16}$ & 4.03 & 34.0$^{\pm 21.0}$ & 950$^{\pm 51}$ &
 4.83  & 1479.3& 7.0$^{\pm 1.9}$  &2.46$^{\pm 0.93}$ \\
3  & 1.25$^{\pm 0.92}$ & 4.83 & 17.7$^{\pm 9.4}$ & 706$^{\pm 71}$ &
 5.64  & 1036.2& 5.0$^{\pm 1.1}$  &0.83$^{\pm 0.34}$ \\
4  & 0.64$^{\pm 0.14}$ & 3.22 & 19.0$^{\pm 7.9}$ & 953$^{\pm 81}$ &
 4.03  & 824.1& 5.3$^{\pm 1.2}$  &1.04$^{\pm 0.44}$ \\
5  & 1.30$^{\pm 1.0}$ & 3.22 & 15.2$^{\pm 8.4}$  & 665$^{\pm 60}$ &
 5.64  & 944.8& 5.2$^{\pm 1.2}$ &0.71$^{\pm 0.31}$ \\
6  & 4.00$^{\pm 3.7}$ & 4.03 & 24.0$^{\pm 12.0}$ & 886$^{\pm 115}$ &
 4.83  & 1119.6& 4.0$^{\pm 0.8}$ &0.23$^{\pm 0.12}$ \\
7  & 2.19$^{\pm 0.47}$ & 3.22 & 58.0$^{\pm 22.0}$& 771$^{\pm 170}$ &
 4.83  & 3109.4& 4.3$^{\pm 0.8}$ &0.33$^{\pm 0.15}$ \\
8  & 1.45$^{\pm 0.33}$ & 2.42 & 16.6$^{\pm 5.1}$ &1146$^{\pm 76}$&
 3.22  & 598.7& 4.3$^{\pm 0.9}$ &0.34$^{\pm 0.18}$ \\
9  & 3.60$^{\pm 2.1}$ & 3.22 & 18.0$^{\pm 8.2}$ & 850$^{\pm 145}$&
 4.03  & 875.3& 3.7$^{\pm 0.6}$ &0.17$^{\pm 0.08}$ \\
10 & 2.40$^{\pm 2.2}$ & 2.42 & 24.0$^{\pm 15.0}$ & 1163$^{\pm 82}$&
 4.83  & 853.0& 5.6$^{\pm 1.7}$ &0.53$^{\pm 0.28}$ \\
\hline\hline
\end{tabular}
\end{flushleft}
     } 
\scriptsize{\begin{itemize}
\item {\bf Parameters of the clusters (columns 2, 3 and 4), radii of
the associated HII regions} \\
{\bf (column 6) and uncertainties in their determination are taken
from Melo et al. (2005). \\
\vspace{-0.7cm}
\item Ionized gas density (column 5) was derived from PMAS observations. \\
\vspace{-0.7cm}
\item One pixel uncertainty ($\pm 0.81$ pc) was adopted in the determination 
of all radii.} 
\end{itemize}}  
\end{table}

\section{Results and discussion}

Each of the selected clusters (see Table 2) is surrounded by a compact HII 
region and has all attributes required by our model. 
We solve Equation (\ref{eq2}) by iteration with the relative accuracy
$\Delta \eta / \eta \le 10^{-5}$. Equations (\ref{eq2a}), (\ref{eq2b}) and 
(\ref{eq3}) were used every time when the iteration procedure requires 
new values for the threshold luminosity, $L_{crit}$, cooling function,
$\Lambda_{st}$, and the star cluster wind terminal speed, $V_{\infty}$.
Our results for each of the considered clusters are shown in Table 2 
(column 9). We use the error propagation equation (Bevington \& Robinson, 
2003) in order to calculate the errors provided by the uncertainties in the 
determination of the input parameters of the model: $M_{SC}$, 
$R_{SC}$, $N^{SC}$, $R_{HII}$ and $n_{HII}$. Unfortunately, the uncertainties 
in the determination of the star cluster radii and sizes of the HII regions 
are not presented in the original paper of Melo et al. (2005). We 
take a 1 pixel ($\pm 0.81$~pc) as a conservative estimate for the uncertainties
in the measured radii. 

The results of the calculations are presented in Figure 3, where the heating
efficiency is presented as a function of star cluster radii and masses.
It seems that there is a trend on panel b for the heating efficiency to be 
larger for less massive clusters. However this must be confirmed with better
sets of input data. We also suggest for a future analysis that the star 
cluster stellar density may be a better input parameter, which combines the 
two major observables, the star cluster mass and radius, into a single 
parameter.  
\begin{figure}[htbp]
\plottwo{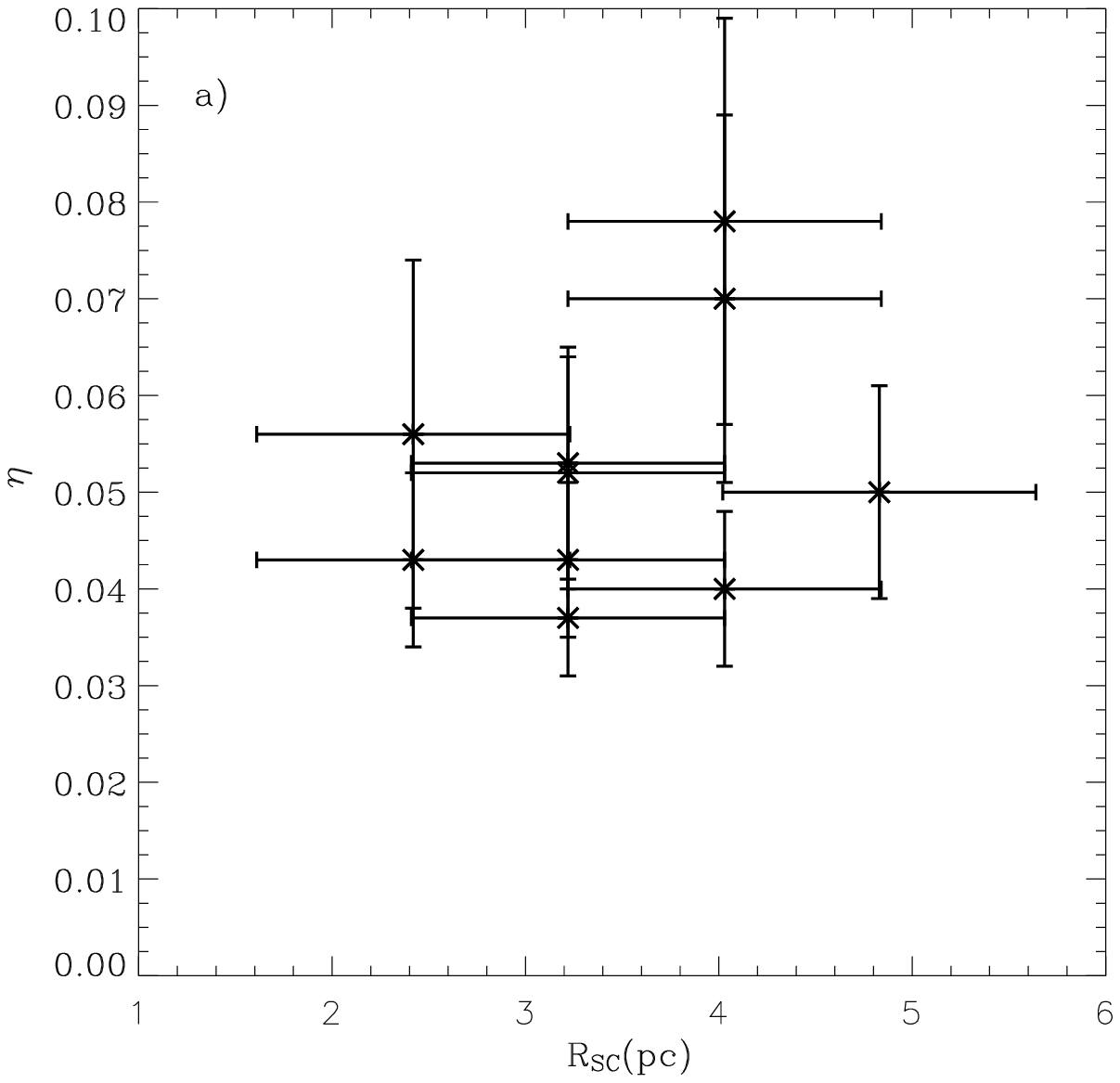}{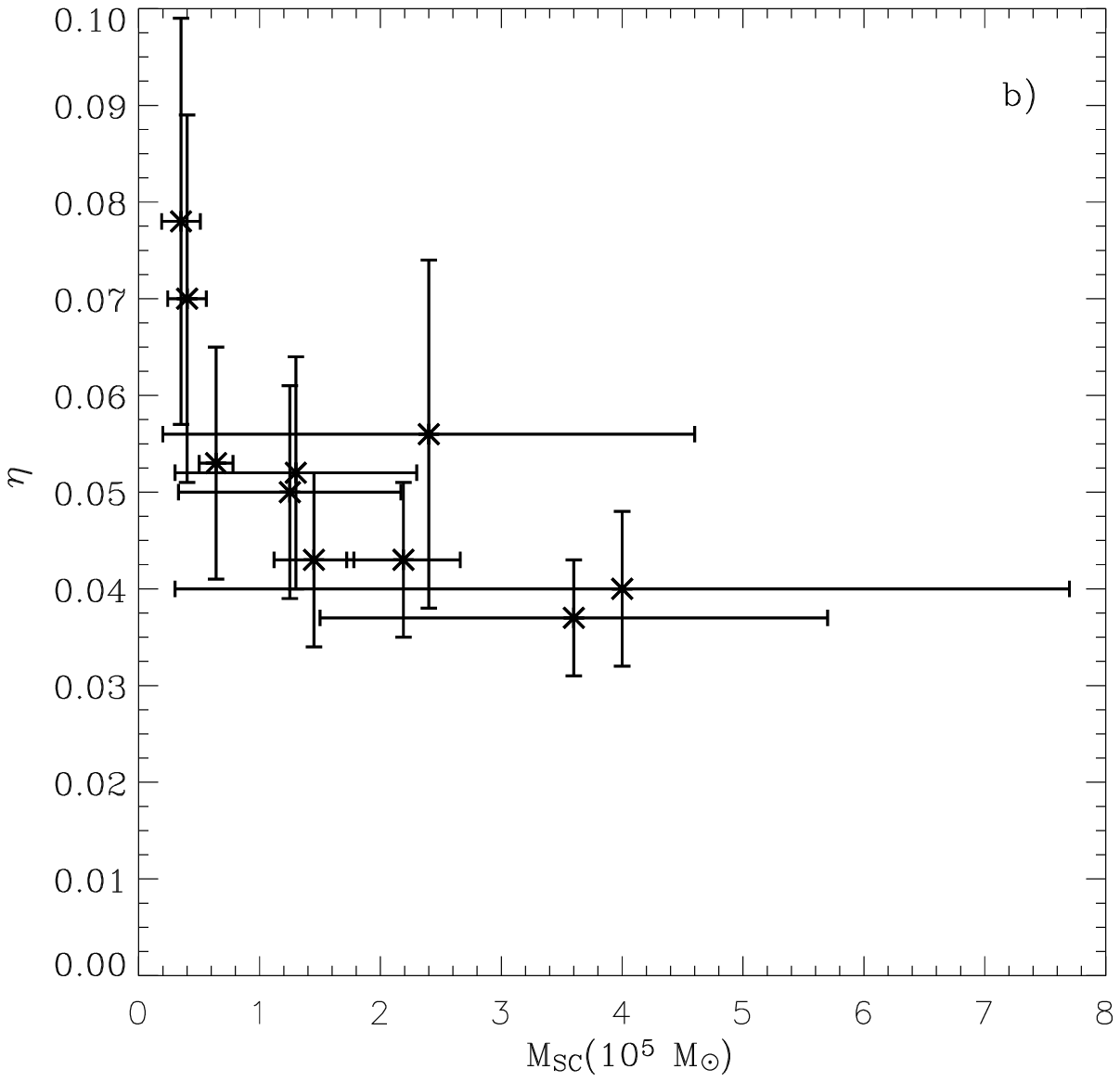}
\caption{The calculated heating efficiency. Panels a and b present 
the star clusters heating efficiency as a function of star cluster 
radius and mass, respectively.}
\label{fig3}
\end{figure}

Figure 3 shows that the heating efficiency does not exceed $10\%$ 
for all clusters in our sample. This implies that our massive and compact 
stellar clusters have a much reduced outflow velocity and negative feedback 
into the ambient ISM than what one would expect using synthetic models. 
Indeed, the mechanical energy output rate is:
\begin{equation}
\label{eq6}
L_{out} = \frac{1}{2} {\dot M}_{out} V^2_{\infty} , 
\end{equation}
where the star cluster wind terminal speed, $V_{\infty}$, is defined by 
equations (\ref{eq3}) and (\ref{eq2b}) and the mass output rate, 
${\dot M}_{out}$, is ( W\"unsch et al. 2007):
\begin{equation}
\label{eq7}
{\dot M_{out}} = {\dot M}_{SC} \left(\frac{L_{crit}}{L_{SC}}\right)^{1/2} . 
\end{equation}
The critical luminosity, $L_{crit}$, is defined by equation (\ref{eq2a}).
The fraction of the injected mechanical energy, $L_{out}/L_{SC}$, that a 
cluster returns to the ambient ISM thus is:
\begin{equation}
\label{eq8}
\frac{L_{out}}{L_{SC}} = \left(\frac{L_{crit}}{L_{SC}}\right)^{1/2} 
                         \left(\frac{V_{\infty}}{V_{A\infty}}\right)^2 . 
\end{equation}

The calculated mechanical energy output does not exceed a few per cent 
of the star cluster mechanical luminosity for all selected clusters (see 
Table 2).
Only in this way the shock-heated matter driven out as a cluster wind can 
cool rapidly to $T \le 10^4$~K and be photoionized while a high ambient 
pressure prevents its expansion into the surrounding interstellar medium. 

The implication of our results, when compared with the recently inferred 
(Strickland \& Heckman, 2009) net efficiency of supernova and stellar wind 
feedback in the nucleous of M82 ($\ge 30$\%),   
is that there is a phase, a time during which 
massive and compact clusters have a low heating efficiency and undergo a 
bimodal hydrodynamic solution returning to the ISM of their host galaxy only 
a small fraction of mass and mechanical energy released inside the star 
cluster volume. Here we suggest that the selected young, massive clusters 
pass through such special phase in their hydrodynamical evolution, 
highlighted observationally by the presence of a compact HII region. 
Indeed, the relevant cluster parameters such as: the energy 
and mass deposition rates, the mean separation between nearby energy sources 
and the chemical composition of the injected matter - all change with time. 
This must lead to important changes in $\eta$ and thus to large displacements
of the threshold luminosity and noticeable changes 
in the rates of mass, $\dot M_{out}$, and energy, $L_{out}$, 
which a star cluster returns to the ISM. The time evolution of $\eta$ 
will be the subject of a forthcoming communication.

\acknowledgments 
We thank our anonymous referee for a critical review. We also appreciate 
fruitful discussions with Divakara Mayya and Daniel  Rosa-Gonz\'alez dealing 
with their selection criteria of stellar clusters in M82. 
This study has been supported by CONACYT - M\'exico, research grants 
82912 and 60333, and partially funded by AYA2007-67965-CO3-O1 from the 
Spanish Consejo Superior de Investigaciones Cient\'\i{}ficas and
the Spanish MEC under the Consolider-Ingenio 2010 Program grant CSD2006-00070:
First Science with the GTC.


\begin{thebibliography}{99}

\bibitem{1} Bevington, P. R. \& Robinson, D. K. {\it Data Reduction and
            Error Analysis for the Physical Sciences}, 2003, The McGraw-Hill
            Companies, NY, p. 41

\bibitem{2} Bisnovatyi-Kogan, G. S. \& Silich, S. A. 1995, Rev. Mod.
            Phys. 67, 661

\bibitem{3} Bradamante F., Matteucci, F. \& D'Ercole, A. 1998, A\&A,
            337, 338
\bibitem{4} Chevalier, R. A. \& Clegg, A. W. 1985, Nature, 317, 44 
\bibitem{5} Chu, Y.-H., Chang, H.-W., Su, Y.-L. \& Mac Low, M.-M. 1995,
            ApJ, 450, 157
\bibitem{6} Cooper, J. L., Bicknell, G. V., Sutherland, R. S. \&
            Bland-Hawthorn, J. 2008,  ApJ, 674, 157
\bibitem{7} Ehlerov\'a, S., Palou\v s, J. \& W\"unsch, R. 2004,
            ApSS, 289, 279
\bibitem{8} H\"agele, G. F., D\'\i az, A. I., Cardaci, M. V., Terlevich, E.
            \& Terlevich, R. 2007, MNRAS, 378, 163 
\bibitem{9} Leitherer, C., Schaerer, D., Goldader, J.D. et al., 
            1999, ApJS, 123, 3 
\bibitem{10} Lozinskaya, T. 1992, Supernovae and stellar wind in the 
             interstellar medium, NY: American Institute of Physics
\bibitem{11} Luo, D., McCray, R. \& Mac Low, M.-M. 1990, ApJ, 362, 267
\bibitem{12} Mac Low, M.-M. \& McCray, R. 1988, ApJ, 324, 776
\bibitem{13} Mayya, Y. D., Romano, R., Rodr\'\i{}guez-Merino, L. H., Luna, A.,
             Carrasco, L. \& Rosa-Gonz\'alez, D. 2008, ApJ, 679, 404 
\bibitem{14} Meaburn, J. 1980, MNRAS, 192, 365 
\bibitem{15} Melo, V. P., Mu\~noz-Tu\~n\'on, C., Ma\'\i{}z-Apell\'aniz, J.
            \& Tenorio-Tagle, G. 2005, ApJ, 619, 270
\bibitem{16} O'Connell, R. W. \& Mangano, J. J. 1978, ApJ, 221, 62
\bibitem{17} Puche, D., Westpfahl, D., Brinks, E. \& Roy, J.-R. 1992,
             AJ, 103, 1841
\bibitem{18} Recchi, S.,  Matteucci, F. \& D'Ercole, A. 2001, MNRAS, 322, 800
\bibitem{19} Roth, M.~M., Kelz, A., Fechner, T. et al. 2005, \pasp, 117, 620
\bibitem{20} Russ, J. C. {\it The image processing handbook}, 2002, CRC Press, 
             p.527 
\bibitem{21} Shaw, R.~A., \& Dufour, R.~J. 1995,  \pasp, 107, 896
\bibitem{22} Silich, S., Tenorio-Tagle G. \& A\~norve Zeferino, G.A.
              2005, ApJ, 635, 1116
\bibitem{23} Silich, S., Tenorio-Tagle, G. \& Mu\~noz-Tu\~n\'on, C. 2007,
             ApJ, 669, 952
\bibitem{24} Silich, S., Tenorio-Tagle G. \& 
             Rodr\'\i{}guez Gonz\'alez, A. 2004, ApJ, 610, 226
\bibitem{25} Smith, L.J., Westmoquette, M.S., Gallagher, J.S. III, 
              O'Connell, R.W., Rosario, D.J. \& de Grijs, R. 2006, 
              MNRAS, 370, 513
\bibitem{26} Stevens, I. R., Blondin, J. M. \& Pollock, A. M. T. 1992,
             ApJ, 386, 265
\bibitem{27} Stevens, I.R. \& Hartwell, J.M. 2003, MNRAS, 339, 280
\bibitem{28} Strickland, D. K. \& Heckman, T. M. 2009, astro-ph/0903.4175
\bibitem{29} Tenorio-Tagle, G., Mu\~noz-Tu\~n\'on, C., P\'erez, E., 
             Silich, S. \& Telles, E. 2006, ApJ, 643, 186
\bibitem{30} Tenorio-Tagle G., Silich, S. \& Mu\~noz-Tu\~n\'on, C.
             2003, ApJ, 597, 279
\bibitem{31} Tenorio-Tagle, G., Silich, S., Rodr\'iguez-Gonz\'alez A. 
             \& Mu\~noz-Tu\~n\'on, C., 2005, ApJ Lett. 628, L13
\bibitem{32} Tenorio-Tagle, G., W\"unsch, R., Silich, S. \& Palou\v{s}, J.
             2007, ApJ, 658, 1196
\bibitem{33} Weaver, R., McCray, R., Castor, J., Shapiro, P. \& 
             Moore, R. 1977, ApJ, 218, 377
\bibitem{34} W\"unsch, R.,  Silich, S. Palou\v{s}, J. \& Tenorio-Tagle, G.
             2007, A\&A, 471, 579
\bibitem{35} W\"unsch, R.,  Tenorio-Tagle, G., Palou\v{s}, J. \& Silich, S.
             2008, ApJ, 683, 683
\end{thebibliography}
\end{document}